\documentclass{article}
\usepackage[utf8]{inputenc}

\usepackage{amssymb}
\usepackage{amsmath}
\usepackage{amsthm}
\usepackage{graphicx}
\usepackage{booktabs}
\usepackage{siunitx}
\usepackage{hyperref}
\usepackage{algorithm}
\usepackage{algpseudocode}

\title{Item Recommendation from Implicit Feedback}

\newcommand{\argmin}{\operatorname*{argmin}}
\newcommand{\argmax}{\operatorname*{argmax}}
\newcommand{\bw}{\mathbf{w}}
\newcommand{\bh}{\mathbf{h}}
\newcommand{\bx}{\mathbf{x}}
\newcommand{\bz}{\mathbf{z}}
\newcommand{\btheta}{\boldsymbol{\theta}}
\newcommand{\sembu}{\phi}
\newcommand{\sembi}{\psi}
\newcommand{\fembu}{\boldsymbol{\phi}}
\newcommand{\fembi}{\boldsymbol{\psi}}
\newcommand{\embu}{\fembu(c)}
\newcommand{\embi}{\fembi(i)}
\newcommand{\embj}{\fembi(j)}

\newcommand{\gradbtheta}{\nabla_{\btheta} }
\newcommand{\gradgradbtheta}{\nabla^2_{\btheta} }

\newcommand{\ttheta}{\tilde{\theta}}
\newcommand{\bttheta}{\boldsymbol{\ttheta}}
\newcommand{\gradttheta}{\nabla_{\ttheta} }
\newcommand{\gradgradttheta}{\nabla^2_{\ttheta} }
\newcommand{\gradbttheta}{\nabla_{\bttheta} }
\newcommand{\gradgradbttheta}{\nabla^2_{\bttheta} }

\newcommand{\hy}{\hat{y}}
\newcommand{\hyc}[1]{\hat{y}(#1|c)}
\newcommand{\hr}{\hat{r}}
\renewcommand{\O}{\mathcal{O}}
\newcommand{\true}{\text{true}}
\newcommand{\grad}{\nabla}
\newcommand{\talpha}{\tilde{\alpha}}
\newcommand{\rightpropto}{\mathbin{\reflectbox{$\propto$}}}

\author{
 Steffen Rendle\thanks{Google Research, Mountain View, USA}\\
  \texttt{srendle@google.com}
}
\date{\vspace{-6.0ex}}

\begin{document}

\maketitle

\begin{abstract}
The task of item recommendation is to select the best items for a user from a large catalogue of items.
Item recommenders are commonly trained from implicit feedback which consists of past actions that are positive only.
Core challenges of item recommendation are
(1)~how to formulate a training objective from implicit feedback and
(2)~how to efficiently train models over a large item catalogue.
This article provides an overview of item recommendation, its unique characteristics and some common approaches.
It starts with an introduction to the problem and discusses different training objectives.
The main body deals with learning algorithms and presents sampling based algorithms for general recommenders and more efficient algorithms for dot product models.
Finally, the application of item recommenders for retrieval tasks is discussed. 
\end{abstract}

\section{Introduction}

Item recommendation, also known as top-n recommendation, is the task to select the best items from a large catalogue of items to a user in a given context.
For example, an e-commerce website recommends products to their customers, or a video platform might want to select  videos that matches a user's interests.
Item recommendation models are usually learned from implicit feedback.
Instead of recommending based on explicit preferences such as \emph{a user gave 4 stars to a movie}, the recommendations are based on the past interactions of the user with the system.
For example, recommendation can be based on past products purchased by a customer, or videos watched by a user.
A good recommender system is often contextual, considering the situation at which a recommendation should be made, e.g., what product is the user currently browsing on, or time information such as weekday vs weekend.

Item recommendation can be formulated as a context-aware ranking problem where the whole set of items should be ordered given a query context.
A scoring function that expresses the preference between a context and an item is used to order the items.
This article covers learned scoring functions that are optimized from implicit feedback.
The core challenges of learning from implicit feedback are to formulate an optimization objective and to design algorithms that can handle large item catalogues.
First, implicit feedback contains weak signals about a user's preference: the selected items in the past give a weak indicator of what a user likes.
Implicit feedback usually does not contain negative interactions, instead weak negatives are derived from all the remaining items, i.e., the items that a user has not interacted with.
Considering all remaining items as weak negative signal leads to the second challenge of high training costs.
Formulating the item recommendation problem as a standard machine learning task is very costly because every positive signal from the implicit feedback entails negative signal over all items.
That makes off-the-shelve training algorithms hard to apply to item recommendation models.
Item recommenders are usually trained either (a)~using sampling or (b)~specialized algorithms taking into account model and loss structure.

This article starts with a detailed discussion of the item recommendation problem, covering the scoring function, the retrieval task, common evaluation metrics, and a summary of the core challenges in learning item recommendation from implicit feedback.
The main part of this article focuses on the training problem.
Sampling methods for general scoring functions based on pointwise, pairwise and softmax losses are described.
The sampling distribution and weighting function are important for fast convergence and for aligning the algorithm better with the evaluation metrics.
Section~\ref{sec:learning_dot_product} discusses more efficient training algorithms that can be applied to dot product models and square losses.
Finally, at application time, item recommenders need to be able to retrieve the highest scoring items quickly from the whole item catalogue.
Section~\ref{sec:serving} discusses sublinear time approaches for the retrieval task.

\section{Problem Definition}

The goal of item recommendation is to retrieve a subset of interesting items for a given context $c \in C$ from a set of items $I$.
In traditional collaborative filtering, the context would be the user, in more complex cases, the context might be user and time, a user and location, the sequence of previously selected items by a user, etc.
To cover these cases, this article uses the general concept of a context as a placeholder.
Also the input representation of a context and an item is flexible, it could be a user id or item id, but also more complex such as image pixels of an item, a textual description, etc.
Such attributes are especially important in the cold start problem. 
The discussion in this article is independent of these choices and should be able to accommodate most item recommendation settings.

\begin{table}
\centering
\caption{Frequently used notation}
\label{tbl:notation}
\begin{tabular}{|l|l|}
    \hline
    Symbol & Description \\
    \hline
    $I$ & set of items \\
    $C$ & set of context \\
    $S \subseteq C \times I$ & set of implicit observations \\
    $I_c$ & set of items selected in context $c$, i.e., $I_c := \{i:(c,i) \in S\}$ \\
    $C_i$ & set of context that selected item $i$, i.e., $C_i :=\{c:(c,i) \in S\}$\\
    $\hyc{i}$ & (learnable) scoring function over context-item pairs \\
    $\btheta$ & model parameters of the scoring function $\hy$ \\ 
    $\embu$ & (learnable) context representation/ embedding \\
    $\embi$ & (learnable) item representation/ embedding \\
    $\langle \cdot, \cdot \rangle$ & inner/ dot/ scalar product between two vectors or two matrices \\
    $\otimes$ & outer product between two vectors \\
    $\delta(b)$ & indicator function, $1$ if $b$ is true, $0$ otherwise\\
    \hline
\end{tabular}
\end{table}

\subsection{Recommender Modeling}

The preference of a context $c \in C$ to an item $i \in I$ is given by a scoring function 
\begin{align}
    \hy(c,i) = \hyc{i} ,\quad \hy : C \times I \rightarrow \mathbb{R}
\end{align}
The choice and design of the scoring function $\hy$ is the problem of recommender system modeling.
Most of the discussion here is independent of the particular choice of $\hy$.

The scoring function is parametrized\footnote{To be precise, $\hy$ should be $\hy_{\btheta}$, but for convenience, the subscript is omitted in this article.} by a set of model parameters $\btheta \in \mathbb{R}^p$.
Learning the model parameters is a core problem within item recommendation and the main focus of this article.
The first derivative of $\hyc{i}$ with respect to a parameter vector $\btheta$ is denoted as $\gradbtheta \hyc{i}$, and the second derivative as $\gradgradbtheta \hyc{i}$.
Table~\ref{tbl:notation} contains an overview of frequently used symbols.

\subsubsection{Dot Product Models}

While most of this article does not make any assumption about the scoring function, many recommender models have additional structure which is discussed now.
For item recommendation, $\hy$ is often decomposable into a dot product between a context and item representation
\begin{align}
    \hyc{i} = \langle \embu, \embi \rangle
\end{align}
where context and items are represented by $d$ dimensional embeddings
\begin{align}
    \fembu : C \rightarrow \mathbb{R}^d, \quad \fembi: I \rightarrow \mathbb{R}^d .
\end{align}
Again, $\fembu$ and $\fembi$ are parametrized by $\btheta$.
Such models are also known as two tower models, or dual encoders~\cite{yang:twotower}. 

Unless stated otherwise, the methods described in this article will be general and are \emph{not} restricted to dot product models.
However, dot product models will be revisited frequently as they have very desirable properties for item recommendation.

\subsubsection{Examples}
Two common examples for scoring functions with a dot product are matrix factorization and two tower deep neural networks. 
First, the scoring function of matrix factorization is $\hyc{i} = \langle \bw_c, \bh_i \rangle$ where each context (or item) is directly represented by a $d$ dimensional embedding vector, $\bw_c \in \mathbb{R}^d$ (or $\bh_i \in \mathbb{R}^d$).
These embedding matrices, $W \in \mathbb{R}^{C \times d}$ and $\ H \in \mathbb{R}^{I \times d}$ are the model parameters $\btheta$.
A matrix factorization is a dot product model with $\embu = \bw_c$ and $\embi = \bh_i$.

Second, a general two tower deep neural network (DNN), $\hyc{i} = \langle \embu, \embi \rangle$, where $\embu$ is a learnable function that extracts a $d$ dimensional representation for each context, and analogously $\embi$ is a learned representation for items.
The model parameters of these functions are $\btheta$.
For example $\embi$ could be a multi layer perceptron that generates an embedding based on the features of an item (e.g., genre, actors, director, release year of a movie) -- here the model parameters $\btheta$ would be the weight matrices of the hidden layers.
Other examples for $\embi$ or $\embu$ would be a convolutional network to extract spatial features from an image and map it to an embedding, or a recurrent model that builds an embedding representation based on sequential data such as previous purchases of a user or textual descriptions of an item.

\subsection{Applying Item Recommenders}

The scoring function $\hy$ can be used to rank items for a given context.
Let $\hr(i|c)$ be the rank/position of item $i$ in the sorted list of items $I$ given context $c$ and scoring function $\hy$,
i.e., $\hr : I \to \{1,\ldots,|I|\}$.
Here $r$ is a bijective mapping -- in case of ties, there is some resolution, e.g., random.
Thus, the inverse function gives the item ranked at a certain position, e.g, $\hr^{-1}(3|c)$ would be the item ranked at the 3rd position for a context $c$.

When a recommendation model $\hy$ is used for the task of item recommendation, it needs to find the $n$ highest scoring items for a given context $c$.
That means it has to compute the items $\hr^{-1}(1|c), \hr^{-1}(2|c), \ldots, \hr^{-1}(n|c)$.
For example, a recommendation platform might want to show the user the 20 items with the highest score.
Section~\ref{sec:serving} will discuss this problem of computing the top $n$ items in more detail.

For a good user experience there are other considerations besides high scores when selecting items.
For example, diversity of results and slate optimization are important factors~\cite{wilhelm:diversity,jiang:slate}.
A naive implementation just returns the top scoring items.
While each item that is shown to a user might be individually a good choice, the combination of items might be suboptimal.
For example, showing item $i$ might make item $j$ less attractive if $i$ and $j$ are very close or even interchangeable.
Instead, it might be better to choose an item $l$ that complements $i$ better -- even if $l$ has a lower score than $j$.
Diversification of result sets is an example to avoid some of these effects.
Commonly, algorithms for slate optimization and diversification are built on top of an item recommender.
The remainder of this article will not discuss these important issues further, instead the article will focus on the top $n$ item recommendation task.

\subsection{Evaluation}
\label{sec:metrics}

The quality of an item recommendation algorithm is typically measured by a ranking metric over the positions at which it ranks relevant items.
For a given context $c$ and a ground truth set of relevant items $\{i_1,i_2,\ldots\}$, the ranks of these relevant items, $R := \{\hr(i_1|c),\hr(i_2|c),\ldots\}$, are computed and then a metric is applied.
Several popular choices are discussed next.

\emph{Precision at position $n$} measures the fraction of relevant items among the top $n$ predicted items:
\begin{align}
    \text{Prec}(R)_n = \frac{|\{r \in R: r \leq n\}|}{n}.
\end{align}
\emph{Recall at position $n$} measures the fraction of all relevant items that were recovered in the top $n$:
\begin{align}
    \text{Recall}(R)_n = \frac{|\{r \in R: r \leq n\}|}{|R|}.
\end{align}
\emph{Average Precision at $n$} measures the precision at all ranks that hold a relevant item:
\begin{align}
    \text{AP}(R)_n = \frac{1}{\min(|R|,n)}\sum_{i=1}^n \delta(i \in R) \text{Prec}(R)_i.
\end{align}
\emph{Normalized discounted cumulative gain (NDCG) at $n$} places an inverse log reward on all positions that hold a relevant item:
\begin{align}
    \text{NDCG}(R)_n = \frac{1}{\sum_{i=1}^{\min(|R|,n)} \frac{1}{\log_2(i+1)}} \sum_{i=1}^n \delta(i \in R) \frac{1}{\log_2(i+1)}.
\end{align}
\emph{Area under the ROC curve (AUC)} measures the likelihood that a random relevant item is ranked higher than a random irrelevant item.
\begin{align}
    \text{AUC}(R) =& \frac{1}{|R|(|I|-|R|)}\sum_{r\in R}\sum_{r' \in (\{1,\ldots, |I|\} \setminus R)} \delta(r < r')
\end{align}
These metrics differ in how much emphasize they place on particular ranks.
Most metrics focus strongly on the top ranks and give almost no rewards for items that appear late in the result list.
AUC is an exception as it has a slow (linear) decay.
A more detailed discussion about these metrics can be found in~\cite{krichene:sampled_metrics}.

The choice of the metric depends on the application.
For retrieval tasks, head-heavy metrics such as Precision, Recall, AP or NDCG are better choices than AUC because users will likely investigate only the highest ranked items.
If the retrieved results are directly shown, NDCG and AP can be better metrics than Precision and Recall because they emphasize the top ranked results stronger, e.g., distinguish between the first and second position.
If the retrieval stage is followed by a reranking step, metrics such as Recall with $n$ equal to the number of candidates that will be reranked can be a better choice than metrics that care about the ranks within the top $n$ such as NDCG or AP.

\subsection{Learning from Implicit Feedback}

This article covers implicit feedback between contexts and items. 
Let $S \subseteq C \times I$ be a set of such observations, where $(c,i) \in S$ is an implicit feedback pair.
Examples for $S$ are clicks on links, video watches, purchases of items.
Implicit observations provide a (weak) positive signal of a user's preferences.
For example, if a user watches a video, it means that this video matches to the user's taste to some degree.
Likewise, the items a user has never considered in the past are indicative of their taste as well.
These non-selected items contain both the items that a user is not interested in and items that should be recommended to the user in the future.
A core problem of learning item recommendation from implicit feedback is to define a training objective for trading off the selected with the non-selected items.
Section~\ref{sec:losses} describes such strategies.
Each of these strategies defines the loss over all selected and non-selected items, i.e., not just over $S$ but over the whole set $C\times I$.
This leads to a second core challenge of deriving efficient learning algorithms.

This differentiates item recommendation from the learning to rank literature~\cite{burges:overview} where a smaller set of candidates needs to be reranked.
This reranking of a small set of candidates has different problem characteristics than item recommendation because (i) it can be done directly on clickthrough data \cite{joachims:searchengine} avoiding the problem to define labels on unobserved data and (ii) it is computationally simpler because it does not involve the full catalogue of all items.
Such models serve a different purpose than the item recommendation models that are discussed in this article, and applying them directly to retrieval over the whole corpus is prone to low retrieval quality due to folding~\cite{xin:folding}.
However, they are very useful as a second stage reranker on top of the retrieval models that are the focus here.
For example, \cite{covington:youtube} describes such a two stage architecture where item recommendation is trained from implicit feedback and a second stage reranks the top $n$ retrieved items based on impression data. 
In this case, the primary goal of item recommendation is to achieve a high recall within the top $n$ results because the final ranking of these candidates is determined by the second stage ranker.

Another class of recommender system problems is based on explicit feedback, most notably \emph{rating prediction}.
Here, explicit preferences are provided such as \emph{a user $u$ assigns 5 stars to a movie $i$, but only 2 stars to movie $j$}.
This is a different problem from item recommendation from implicit feedback.
The training objective is simpler because it can be defined directly on the explicit feedback, avoiding the costly optimization over all items.
These aspects makes it similar to the learning to rank problem, however it differs from learning to rank in the metrics.
A commonality of item recommendation from implicit feedback and recommenders over explicit feedback, is that they share similar design patterns for the scoring function.
For example, matrix factorization or item-based collaborative filtering is popular for both problems.

\subsection{Core Challenges of Item Recommendation}
\label{sec:problem_characteristics}

To summarize, the core challenges of item recommendation are:
\begin{enumerate}
    \item Formulating a loss function over implicit feedback. This requires to define a strategy to trade off selected with unselected items while considering the ranking metric.
    \item Learning algorithms that can handle the large number of unselected items efficiently.
\end{enumerate}

\section{Learning Objectives}
\label{sec:losses}

This section describes three popular approaches for casting the item recommendation problem into a supervised machine learning problem.

\subsection{Pointwise Loss}

The observations, $S$, can be directly casted into a binary classification problem by treating the observed feedback $S$ as positive instances and all non selected items, $(C \times I) \setminus S$ as weak negatives.
Each training instance has a positive or negative label and an example weight.
Any positive pair $(c,i) \in S$, is assigned a positive label, e.g., $y=1$ or, if available, another positive signal such as the strength of the interaction, e.g., number of times selected, or time spent.
Additionally all non-observed items $j$ for the context are included with a negative label, e.g., $y=0$ (for regression) or $y=-1$ (for classification).
A weight $\alpha$ can be assigned to each context-item tuple to indicate the confidence about the training case.
Usually, the confidence in the non-observed pairs is lower than for observed one.
That balances the large amount of $\O(|C\times I|)$ negatives with the smaller amount of $|S|$ positives.
Other weighting schemes can be used to downweight popular items to promote more tail items in the result, or reduce the weight of users with many positive interactions so that the experience is not dominated by the behavior of heavy users.

From a regression point of view, a good scoring model predicts scores close to $y$.
From a probabilistic point of view, the binary event $p(i=\true|c)$ is modeled through $\hy$, e.g., $p(i=\true|c) = \sigma(\hyc{i})$.
These ideas can be formalized as a pointwise loss
\begin{align}
    L(\btheta, S) 
     :=& \sum_{c \in C} \sum_{i \in I} \alpha(c,i) \, l(\hyc{i}, y(c,i)) + \lambda(\btheta) \label{eq:ew}
\end{align}
where $\alpha: C \times I \rightarrow \mathbb{R}^+$ is a weight function, $y$ the label function (e.g., $y(c,i) = \delta((c,i) \in S)$), and $\lambda$ is some regularization function and $l$ is a loss function, e.g.,
\begin{subequations}
\label{eqs:losses}
\begin{align}
    l^{\text{square}}(\hy, y) &= (\hy - y)^2 \label{eq:loss_square}\\
    l^{\text{logistic}}(\hy, y) &= \ln(1+\exp(-y\,\hy)) \label{eq:loss_logistic} \\ 
    l^{\text{hinge}}(\hy, y) &= \max(0, 1-y\,\hy) 
    \label{eq:loss_hinge} 
\end{align}
\end{subequations}
This casts the problem of item recommendations to a standard binary classification or regression problem.
What makes this problem special is that the number of training examples is huge $\O(|C| \, |I|)$.
The number of items can be in the millions for large scale problems which makes application of standard learning methods challenging.

Examples for recommender system algorithms that use a pointwise loss are iALS~\cite{hu:ials}, SLIM~\cite{ning:slim}, or iCD~\cite{bayer:icd}.

\subsection{Pairwise Loss}

For retrieval, it is not of primary interest if the score $\hy$ is exactly 1 or 0.
Instead the relative ordering of the items in a context is of central importance.
Given a context $c$, pairwise methods compare all pairs of items $(i,j)$ where $(c,i) \in S, (c,j) \not\in S$.
A successful scoring function, $\hy$, assigns a higher score to the observed item $i$ than the unobserved one: $\hyc{i} > \hyc{j}$.
For the final retrieval, the items are simply sorted by the score.

A general pairwise loss function can be defined as:
\begin{align}
    L(\btheta, S) &:= \sum_{(c,i) \in S} \sum_{j \in I} \alpha(c,i,j)\, l(\hyc{i} - \hyc{j}, 1) + \lambda(\btheta) \label{eq:pw_loss}
\end{align}
where $\alpha$ is a weight function.
See eq~\eqref{eqs:losses} for some example losses $l$.

The number of pairs in this loss is $\mathcal{O}(|S| \, |I|) \supseteq  \mathcal{O}(|C| \, |I|)$.
That means it is at least as expensive as pointwise training, and it faces the same computational challenges.

\subsection{Softmax Loss}

From a probabilistic point of view, pointwise training models the binary event $p(i=\text{true} | c)$ through $\hy$, e.g., $p(i=\text{true} | c) = \sigma(\hyc{i})$.
Pairwise training models the conditional probability of the binary comparison $p(i > j|c)$ through $\hy$, e.g., $p(i > j|c) = p(\hyc{i} > \hyc{j}) = \sigma(\hyc{i} - \hyc{j})$.
A third option is to model the conditional multinomial event $p(i|c)$, i.e., $\sum_{j \in I} p(j|c) = 1$ and $\forall j: 0\leq p(j|c) \leq 1$.
The most common function to translate a real valued score, such as $\hy$, to a multinomial distribution is the softmax:
\begin{align}
    p(i|c) = \frac{ \exp(\nu\, \hyc{i}) } {\sum_{j \in I} \exp(\nu\, \hyc{j})} \propto \exp(\nu\, \hyc{i}).
\end{align}
The numerator of the softmax is often referred to as the \emph{partition function}. 
$\nu \in \mathbb{R}^+$ of the softmax is an (inverse) temperature parameter.
Small choices of $\nu$ (=high temperature) make the distribution more uniform.
Larger choices (=low temperature) make it more spiky and concentrated around the large values.
For $\nu \rightarrow \infty$, softmax approaches the \emph{maximum} operator.
And for this limit the probability is concentrated at one index
\begin{align}
    p(i|c) = \delta\left(i = \argmax_j \hyc{j}\right)
\end{align}
This aligns well with ranking metrics that focus on the top ranked positions.
Softmax is widely used in multiclass classification problems, including natural language models where the classes are words, or image classification where the labels are categories.

To keep the notation short, the remainder of this article ignores $\nu$ because usually $\hy$ can absorb any scaling effect of $\nu$ by increasing the norm of $\hy$.
From the softmax definition of the multinomial probability follows the loss function through the negative log likelihood:
\begin{align}
    L(\btheta,S) =& - \sum_{(c,i) \in S}  \ln p(i|c) + \lambda(\btheta) \notag\\
    =& - \sum_{(c,i) \in S}  \ln \frac{\exp(\hyc{i})}{\sum_{j \in I}\exp(\hyc{j})} + \lambda(\btheta) \notag \\
    =&  - \sum_{(c,i) \in S} \left[ \hyc{i} - \ln \sum_{j \in I}\exp(\hyc{j})\right] + \lambda(\btheta) \label{eq:softmax_loss}
\end{align}
Again, the loss contains a double sum over $S$ and $I$, so the number of elements is $\O(|S|\,|I|)$, or $\O(|C|\, |I|)$ if the partition function is computed once per context.

\section{Sampling Based Learning Algorithms}

Algorithms based on sampling are the most common approaches for learning recommender systems from implicit feedback.
Their basic idea is to iterate over the positive examples $(c,i) \in S$, sample negatives $j$ with respect to a sampling distribution $q(j|c)$ and perform a weighted gradient step. 
This section discusses algorithms for pointwise, pairwise and softmax losses.
The choice of $q$ together with a weighting function $\talpha$ are key for designing the learning algorithms.
These two choices are related to $\alpha$ in the loss formulation.
Many different instances of these algorithms have been proposed by varying $q$ and $\talpha$ and some popular choices are discussed in this section.
Most of the results of this section apply to any scoring function, with the exception of Sections~\ref{sec:impr_bpr} and \ref{sec:kernel_softmax} that are restricted to dot-product models.

\subsection{Algorithms for Pointwise Loss}

The pointwise loss in eq.~\eqref{eq:ew} is defined over all context-item combinations, $C \times I$.
The elements in this set are either a positive observation, if $(c,i) \in S$ or an implicit negative one otherwise.
The positive observations are usually informative while the negative provide weaker feedback and are assigned a smaller weight so that they contribute less to the loss.
A sampling based learning algorithm iterates over the positive observations and samples $m$ negatives.
Algorithm~\ref{alg:pointwise_sgd} sketches this idea.
\begin{algorithm}[t]
  \caption{Pointwise SGD}
  \label{alg:pointwise_sgd}
  \begin{algorithmic}[1]
        \Repeat
            \State sample $(c,i) \in S$ \Comment{positive item}
            \State $\btheta \leftarrow \btheta - \eta\, \left[ \talpha(c,i)\, \gradbtheta l(\hat{y}(c,i),1) +  \gradbtheta \lambda(\btheta) \right]$
            \For{$l \in \{1,\ldots,m\}$} \Comment{$m$ negative items}
                \State sample $j$ from $q(j|c)$
                \State $\btheta \leftarrow \btheta - \eta\, \left[ \talpha(c,j)\, \gradbtheta  l(\hat{y}(c,j),0) + \gradbtheta \lambda(\btheta) \right]$
            \EndFor
        \Until converged
  \end{algorithmic}
\end{algorithm}

The main design choices of the algorithm are the sampling distribution $q$ and the weight $\talpha$.
Their choice is related to $\alpha$ from the loss.
If $q$ and $\talpha$ are independent of the model parameters $\btheta$, then the relationship between the three values is
\begin{align}
    \alpha(c,j) = \begin{cases}
      \talpha(c,j), & (c,j) \in S \\
      m |I_c| q(j|c) \talpha(c,j), & (c,j) \not \in S
    \end{cases}
\end{align}
where the sampling probability $q(j|c) = 0$ for $(c,j) \in S$.
This shows how a weighted loss can be either achieved by changing $q$ or $\talpha$.
For example, if the sampling probability $q$ is uniform
\begin{align}
    \alpha(c,j) = \begin{cases}
      \talpha(c,j), & (c,j) \in S \\
      m \talpha(c,j), & (c,j) \not \in S
    \end{cases}
\end{align}
A non-uniform sampling might be beneficial because most of the items are trivially negative, e.g., it is unlikely that a customer likes a random product from a large catalogue.
Such samples are trivial for the model to predict and presenting an already correctly classified item during training has a (near) zero gradient, so overall convergence will be slow.
Sampling items that have a higher chance to be considered by a user are more informative during learning.

Besides uniform sampling, popularity based sampling is another common approach.
Here, the items are sampled proportional to their empirical frequency in the dataset $q(j|c) \propto |C_j|^\beta$ with a squashing exponent $\beta$ to desharpen the distribution.
Section~\ref{sec:alg_pairwise} describes some more sophisticated sampling algorithms.

\subsubsection{Batching}

The SGD algorithm presented so far performs one update step per context-item pair.
Each step is computationally cheap.
Modern hardware can benefit from larger units of computations, so grouping several training examples can result in more efficient utilization of hardware.
For the proposed SGD algorithm, several positive and negative items can be sampled and grouped into a batch of training examples, and then their updates are performed in parallel.
This scheme is especially useful if the same set of negatives is shared in the batch -- a context independent sampler, e.g., frequency based, naturally fits here.
Even more, instead of sampling from a global distribution, it is also common to use the positive items in the batch as the negatives.
For example, if the batch contains positive observations $(c_1,i_1),(c_2,i_2),\ldots$, then $\{i_1, i_2, \ldots\}$ are treated as the negatives for this batch, e.g., for $c_1$, the negatives would be $(c_1, i_2), (c_1, i_3), \ldots$.
In expectation, this corresponds to choosing the empirical frequency sampler for $q$.
To summarize, batching does not improve the computational costs from a complexity perspective but it lowers the wall time on modern hardware.

\subsubsection{Omitted Details}

For simplicity, algorithm \ref{alg:pointwise_sgd} was sketched with a constant learning rate, $\eta$, but more complex schedules can be applied as well.
Also depending on the form of $\hy$, it is common to update only the subset of model parameters that affects $\hyc{i}$ for the particular choice of $(c,i)$.
For example, in an embedding model, usually only a subset of the embeddings are involved in scoring a pair $(c,i)$, and it is common to apply the gradient step only to this subset of embeddings involved.
These details are omitted because they are orthogonal to the problem of item recommendation from implicit feedback.

\subsection{Algorithms for Pairwise Loss}
\label{sec:alg_pairwise}

The pairwise loss (eq.~\ref{eq:pw_loss}) is a double sum over elements $(c,i) \in S$ and items $j \in I$.
This scheme can be directly used for sampling and gradient steps can be performed for each item-pair $(i,j)$.
Algorithm \ref{alg:pairwise_sgd} shows a SGD algorithm for optimizing a pairwise loss.
\begin{algorithm}[t]
  \caption{Pairwise SGD}
  \label{alg:pairwise_sgd}
  \begin{algorithmic}[1]
        \Repeat
            \State sample $(c,i) \in S$
            \State sample $j$ from $q(j|c,i)$
            \State $\btheta \leftarrow \btheta - \eta   \left[\talpha(c,i,j) \gradbtheta l(\hat{y}(c,i)-\hat{y}(c,j), 1) +  \gradbtheta\lambda(\btheta) \right]$
        \Until converged
  \end{algorithmic}
\end{algorithm}
The design choices of this algorithm are the sampling distribution of negatives, $q(j|c,i)$, and the weight of the gradient $\talpha(c,i,j)$.
These two choices are related to the weight $\alpha(c,i,j)$ in the loss (eq.~\ref{eq:pw_loss}).
Similar to the pointwise loss, if $q(j|c,i)$ and $\talpha(c,i,j)$ are independent of the model parameters then choosing $q$ and $\talpha$ implies a pairwise loss with $\alpha(c,i,j) = q(c,i,j)\,\talpha(c,i,j)$.
Again, fixing two of the quantities will determine the third.
For pairwise optimization, several variations with model dependent sampling and weighting have been proposed.

The next subsections discuss several popular variations of the algorithm.
The first one samples items uniformly and has no weights.
The second scheme adds a weight to consider head-heavy metrics.
The third algorithm samples uniformly from items close or above the positive item, the sampling scheme is used to estimate the rank which is used as a weight.
Finally, a sampler that oversamples items based on their rank is presented.

\subsubsection{Uniform Sampling without Weight}
\label{sec:bpr}

The most simple instance of this algorithm uses a constant weight $\tilde{\alpha}(c,i,j) = 1$ and a uniform sampling probability $q(j|c)\propto 1$, which optimizes an unweighted loss with $\alpha(c,i,j) = 1$.
This loss optimizes the area under the ROC curve, AUC, or equivalently, the likelihood that a random relevant item is ranked correctly above a random irrelevant item.
This loss and algorithm is often referred to as \emph{Bayesian Personalized Ranking (BPR)}~\cite{rendle:bpr}.

There are two major issues with this approach:
\begin{enumerate}
    \item Uniform sampling will sample irrelevant items that can be easily distinguished by the model and thus the gradient of such pairs becomes 0. The progress of the algorithm will be very slow because most of the pairs are not helpful.
    \item Metrics that focus on the top items are often preferred for evaluating item recommendation. As discussed in Subsection~\ref{sec:metrics}, AUC is not a top-heavy metric.
For example, for AUC, the benefit for moving a relevant item from position 1000 to 991 is the same as moving on item from position 10 to 1.
For a metric such as NDCG, the second change is much more beneficial and gives a much larger improvement in the metric.
In most applications, the second change would be considered more important.
\end{enumerate}
Several improvements are discussed next.

\subsubsection{Uniform Sampling with Weights}

A popular approach in the learning to rank community is LambdaRank~\cite{burges:overview} which weights the gradient by its influence on a ranking metric.
Let $M$ be a metric as defined in Section~\ref{sec:metrics}, then the weight of a pair of items $i,j$ given a context $c$ is $\talpha(c,i,j) = |M(\hr(i|c)) - M(\hr(j|c))|$.
That means the weight is directly tied to the metric because it measures the influence of flipping the order of the two items.
A practical difficulty of this approach is that the rank of the items $i$ and $j$ needs to be computed.
For item recommendation with a large number of items this becomes very costly.
This is less of a problem for learning to rank problems where only a small set of items are reranked.
An adaption of LambdaRank to item recommendation was proposed in~\cite{yuan:lambdafm}.

\subsubsection{Adaptive Sampling with Weights}

WARP (Weighted Approximate-Rank Pairwise)~\cite{weston:wsabie} is a pairwise algorithm specifically designed for item recommendation.
The motivation of the WARP loss is to directly learn a ranking metric.
Each rank is associated with a penalty, $\gamma$, where the better the predicted rank of the relevant item, the lower the penalty.
This penalty can be chosen to align with the ranking metric~\cite{usunier:owpc}.
For example, the penalty can be linearly increasing (as in AUC) or be a step function for a metric such as precision@k, or increase by the reciprocal of the position.

WARP uses an online algorithm for optimizing this loss.
First, the sampling distribution is non-uniform.
The algorithm tries to sample an irrelevant item $j$ such that it is ranked close to or above the relevant item, i.e.,  $\hyc{j} + 1 > \hyc{i}$.
This choice is aligned with the hinge loss (eq.~\ref{eq:loss_hinge}), where the gradient is zero if the item is correctly ranked above the margin.
The authors propose a rejection sampling algorithm that draws $j$ uniformly, and if it does not meet the condition, it discards $j$ and samples another item.
This means the sampling distribution is $q(j|i,c) \propto \delta(\hyc{j} + 1 > \hyc{i})$.

Sampling an item $j$ that fulfills the requirement  $\hyc{j} + 1 > \hyc{i}$ becomes increasingly hard, the better the model is and the more items are in the catalogue.
The authors propose to stop the sampling algorithm after $|I|$ repetitions. 
The number of rounds until the algorithm terminates can be used to estimate the rank of the relevant item: $r(i|c) \approx \left\lfloor(\#\text{rounds}-1)/|I|\right\rfloor$.
The penalty, $\gamma$, of the estimated rank is used as the weight in the SGD algorithm:
\begin{align}
    \talpha(c,i,j) = \gamma(\left\lfloor(\#\text{rounds}-1)/|I|\right\rfloor)
\end{align}

The definition of $q$ avoids the problem of sampling pairs that have no influence on the gradient step.
However, while the algorithm avoids steps on non-influentual items, the cost for finding a relevant item are high because it needs to score all discarded items.
Usually, scoring an item and updating an item have similar computational complexity.
This gives WARP at most a constant speedup over a trivial method that does not discard items.
However, sampling is only one part of the WARP algorithm and the more important part is that WARP uses the estimated rank in the weight for the gradient.
This way, WARP optimizes a loss that is more reflective of ranking metrics than uniformly weighted methods such as BPR.

\subsubsection{Adaptive Sampling without Weights}
\label{sec:impr_bpr}

As an improvement of the BPR algorithm, \cite{rendle:bprimpr} propose to sample items based on their rank, here the sampling distribution is $q(j|c) \propto \exp(- r(j|c)/\gamma)$, while keeping the weight constant $\tilde{\alpha} = 1$.
While LambdaRank and WARP introduce a weight on the rank by $\talpha$, here the weight is implicitly introduced through $q$ that depends on the rank.
Note that this sampling distribution depends on the context and the current choice of model parameters.
As the model learns, the sampling distribution adapts. 

The novelty of this work is an algorithm to approximately sample from $q(j|c)$ in constant time.
Unlike all the methods that have been discussed so far and that were applicable to any scoring function, this work assumes a dot product model $\langle \embu, \embi \rangle$.
Next, the sampling algorithm is described briefly and more background can be found in~\cite{rendle:bprimpr}.

For a given context, it first samples an embedding dimension $f^*$ according to
\begin{align}
    q(f|c) = |\sembu_{c,f}| \sigma_f, \quad \sigma_f^2 = \text{Var}(\sembi_{\cdot,f})
\end{align}
The motivation for this choice is that for a given context embedding $\embu$ different embedding dimensions $f \in \{1,\ldots,d\}$ are expected to have different contributions on the score and consequently on the rank.
The larger an entry $|\sembu_{c,f}|$ of the context embedding, the more influential is this embedding dimension.
The sign is ignored at this step because without looking at the item embeddings it is meaningless.
Each dimension is normalized by the variance of the item embeddings to ensure that dimensions with larger entries are overweighted in the sampling step.

Then an item is sampled according to the rank induced by $\embi_{\cdot, f^*}$.
This is done by first sampling a desired rank $r^*$ from $\exp(-r/\gamma)$, this step is independent of the scoring function.
Then the $r^*$-largest item from $\embi_{\cdot, f}$ is returned -- or if the sign of $\sembu_{c,f^*}$ is negative, the $r^*$-smallest (=most negative).
This last step can be achieved by storing a sorted list of items for each embedding dimension.
This sorted list is independent of the context and can be refreshed occasionally.
In total, this is an amortized constant time algorithm for sampling an item approximately from $q(j|c) \propto \exp(- r(j|c)/\gamma)$.

\subsection{Algorithms for Sampled Softmax}

A common strategy for training a model over a softmax loss is to sample $m$ negatives $\{j_1, \ldots, j_m\}$ from a distribution $q(j|c)$ and to compute the partition function on this smaller sample.
A key difference of sampled softmax to negative sampling for pointwise and pairwise losses is that for softmax the sample is applied inside a logarithm (see eq.~\eqref{eq:softmax_loss}).
That makes it difficult to get unbiased estimates.
Sampled softmax is commonly used with a correction to the scores and for a sample $\{j_1, \ldots, j_m\}$ of $m$ negative items, and for an observation, $(c,i) \in S$, a gradient step with respect to the following loss is taken:
\begin{align}
    - \hyc{i} + \ln \left(\exp(\hyc{i}) + \sum_{l=1}^m\frac{\exp(\hyc{j_l})}{m\, q(j_l|c)}\right)
\end{align}
The correction $\frac{1}{m\,q(j|c)}$ ensures that if $m \rightarrow \infty$, then sampled softmax is unbiased~\cite{bengio:sampledsoftmax}.
However, typically a small (finite) set of negatives is sampled and thus sampled softmax is biased.
This means no matter how many training epochs the algorithm is run, as long as $m$ is constant, sampled softmax will converge to a different solution than full softmax.
It was shown that the only way to avoid this issue is to use $q(j|c) = p(j|c)$, i.e., to use the softmax distribution itself for sampling~\cite{blanc:adaptive}.
In this case, sampled softmax is unbiased for any sample size $m$.
For sure, sampling from $p$ is expensive and not a viable option.
Nevertheless, it shows that the bias can be reduced by either increasing $m$ and/or using a sampling distribution closer to $p(j|c)$.

In practice, when optimizing a recommender system with sampled softmax, the sample size $m$ is an important hyperparameter because it directly impacts what loss is optimized.
Typically, a large sample size, $m$, is chosen to mitigate the bias.
Again, the most common sampling distributions are (squashed) popularity (="unigram") sampling or in-batch sampling (=popularity sampling where $m=\text{batchsize}$).
Two more sophisticated sampling approaches that have been specifically designed for softmax are discussed in subsections~\ref{sec:kernel_softmax} and \ref{sec:tapas}.

Algorithm~\ref{alg:softmax_sgd} sketches pseudo code for optimizing a model with SGD for sampled softmax.

\begin{algorithm}[t]
  \caption{Sampled Softmax SGD}
  \label{alg:softmax_sgd}
  \begin{algorithmic}[1]
        \Repeat
            \State sample $(c,i)$ from $S$
            \State sample $\{j_1, \ldots, j_m\}$ from $q(j|c)$
            \State $\btheta \leftarrow \btheta - \eta \gradbtheta\left[-\hyc{i} +
             \ln \left(\exp(\hyc{i}) + \sum_{l=1}^m\frac{\exp(\hyc{j_l})}{m\, q(j_l|c)}\right)  + \lambda(\btheta) \right]$
        \Until converged
  \end{algorithmic}
\end{algorithm}

\subsubsection{Kernel based Sampling}
\label{sec:kernel_softmax}

As discussed before, the sampling distribution is very important for sampled softmax optimization because it lowers the number of samples, $m$, required while keeping the bias low.
The ideal sampling distribution would be $p(j|c,\btheta) \propto \exp(\hyc{j})$ -- or for dot product models $p(j|c,\btheta) \propto \exp(\langle \embu, \embj \rangle)$.
This formulation can be seen as a kernel and a mapping function $\pi: \mathbb{R}^d \rightarrow \mathbb{R}^D$ from the original embedding space to a larger space can be introduced, such that
\begin{align}
    p(j|c,\btheta) \propto \exp(\langle \embu, \embj \rangle \approx \langle \pi(\embu), \pi(\embj) \rangle \rightpropto q(j|c,\btheta)
\end{align}
Such a decomposition allows to compute the (approximated) partition function as:
\begin{align*}
    \sum_{j \in I} \exp(\hyc{j}) \approx \sum_{j \in I}\langle \pi(\embu), \pi(\embj) \rangle = \left\langle \pi(\embu), \underbrace{\sum_{j \in I}\pi(\embj)}_{\bz \in \mathbb{R}^D} \right\rangle = \left\langle \pi(\embu), \bz  \right\rangle
\end{align*}
The key here is that $\bz$ is independent of the context and can be precomputed.
Then the sampling probability of an element j, $q(j|c)$, can be computed in $\O(D)$ time.
Based on this observation, a divide and conquer algorithm can be derived to sample j from $q(j|c, \btheta)$ in $\O(D\,\log_2 n)$ time (see \cite{blanc:adaptive} for details).
Quadratic expansion~\cite{blanc:adaptive} and random feature maps \cite{rawat:rffsoftmax} have been explored for $\pi$.

\subsubsection{Two-pass Sampler}
\label{sec:tapas}

Another sampling strategy is a two stage sampler~\cite{bai:tapas}, where first a large set of $M>m$ items is sampled and only the $m$ most promising candidates are accepted.
The first set is sampled from the squashed empirical frequency distribution, typically $1\%$ to $10\%$ of the items $I$ are sampled.
Then all items on this large set are scored against a batch of context and a smaller set of the $m < M$ highest scoring items is returned.
For efficient computation, it was proposed to distribute the computation of the scores of the  $M$ items over several machines, preferably machines with GPUs.
Finally, sampled softmax is applied on the smaller set.
More details can be found in~\cite{bai:tapas}.

\subsubsection{Relation of Sampled Softmax To Pairwise Loss}

Finally, a relationship between sampled softmax and a pairwise loss is highlighted. 
When the number of negative samples in sampled softmax is $m=1$, then
\begin{align*}
    - \hyc{i} + \ln \left(\exp(\hyc{i}) + \frac{\exp(\hyc{j})}{q(j|c)}\right) 
    =& \ln\frac{\exp(\hyc{i}) + \frac{\exp(\hyc{j})}{q(j|c)}}{\exp(\hyc{i})}\\
    =& l^{\text{logistic}}\left(\hyc{i} - \hyc{j} + \ln q(j|c), 1\right)
\end{align*}
Which is a pairwise logistic loss where the prediction of the negative item is shifted by the correction term $\ln q(j|c)$.

\section{Efficient Learning Algorithms For Special Cases}
\label{sec:learning_decomp}
\label{sec:learning_dot_product}

This section investigates efficient learning algorithms for dot product models, i.e., $\hyc{i} = \langle \embu, \embi \rangle$, and square losses.
First efficient alternating least squares and gradient descent learning algorithms are presented for pointwise losses.
Finally, their application to pairwise square losses is discussed briefly.  

\subsection{Pointwise Square Loss}

The following restrictions are made: (1)~the weights and labels for all unobserved tuples $(c,i) \not\in S$ is constant: let their weight be $\alpha_0$ and the label $0$.
It follows
\begin{align*}
     L(\btheta, S) = &\sum_{c \in C} \sum_{i \in I} \alpha(c,i) (\hyc{i} - y(c,i))^2 + \lambda(\btheta) \\
    =&\sum_{(c,i) \in S} [\alpha(c,i) (\hyc{i} - y(c,i))^2 - \alpha_0 \hyc{i}^2] + \alpha_0 \sum_{c \in C} \sum_{i \in I} \hyc{i}^2 + \lambda(\btheta)
\end{align*}
This can be further simplified to
\begin{align}
    \tilde{L}(\btheta, \tilde{S}) = \sum_{(c,i,\alpha,y) \in \tilde{S}} \alpha (\hyc{i} - y)^2 + \alpha_0 \sum_{c \in C} \sum_{i \in I} \hyc{i}^2 + \lambda(\btheta) 
\end{align}
With $\tilde{S} = \{(c,i,\alpha(c,i) - \alpha_0, y \alpha(c,i)/(\alpha(c,i) - \alpha_0)) : (c,i) \in S\}$ and both $\tilde{L}$ and L share the same optimum (see~\cite{bayer:icd} for more details).
The remainder of this section uses this simplified formulation.

\subsubsection{Gramian Trick}

The first part of the loss $\tilde{L}$ depends only on the small set of observed positive data and is cheap to compute.
The second part appears to be expensive with $|C||I|$ terms and a naive computation is prohibitively costly.
However, the \emph{Gramian trick} makes the computation of the second part very efficient~\cite{bayer:icd}:
\begin{align}
    \alpha_0 \sum_{c \in C} \sum_{i \in I} \hyc{i}^2 =&\alpha_0 \sum_{c \in C} \sum_{i \in I}  \langle \embu, \embi \rangle^2  \\
    =&\alpha_0 \left\langle \sum_{c \in C} \embu \otimes \embu, \sum_{i \in I} \embi \otimes \embi \right\rangle \\
    =&\alpha_0 \left\langle G^C, G^I \right\rangle
\end{align}
with Gram matrices
\begin{align}
    G^C = \sum_{c \in C} \embu \otimes \embu, \quad G^I = \sum_{i \in I} \embi \otimes \embi \label{eq:gramians}
\end{align}
The advantage of this \emph{Gramian trick} is that the sum over all context-item combination is simplified as a  the dot product over two Gram matrices, where each matrix has size $d \times d$.
The cost for computing the Gramians\footnote{The analysis here ignores the cost for computing the embeddings $\embu$ and $\embi$. The derived results have a linear complexity in the costs for computing the embeddings.} is $\O(d^2 (|I| + |C|))$.
The cost for computing the loss, $\tilde{L}$, assuming the Gramians are known is $\O(|S|\,d + d^2)$.
The overall costs are dominated by the loss over the positive examples, $|S|$, as long as $d \leq |S|/(|C|+|I|)$.
This means the loss over \emph{all} examples (including the implicit negative ones) can be computed without paying the computational costs for the negative ones.

The remainder of this section derives efficient solvers for the loss with the Gramian formulation 
\begin{align}
    \tilde{L}(\btheta,\tilde{S}) = \sum_{(c,i,\alpha,y) \in \tilde{S}} \alpha (\hyc{i} - y)^2 + \alpha_0 \left\langle G^C, G^I \right\rangle + \lambda(\btheta) \label{eq:gramloss}
\end{align}
The Gramian trick is related to the Kernel softmax (see Section~\ref{sec:kernel_softmax}) where $\exp(\langle\embu,\embi\rangle) \approx \langle \pi(\embu), \pi(\embi)\rangle$.
In particular, using the square function instead of the exp, $\langle\embu,\embi\rangle^2 = \langle \pi(\embu), \pi(\embi)\rangle$ for $\pi(\bx)=\bx \otimes \bx$. 

\subsubsection{Coordinate Descent/ ALS Solver for Multilinear Models}

The Gramian trick can be used to derive efficient alternating least squares solvers for the family of multilinear models.
This section starts with a recap of multilinear models and then derives a learning algorithm.
Besides multilinearity, it is assumed that the regularization $\lambda(\btheta)$ is a L2 regularization, i.e., $\lambda(\btheta) = \lambda ||\btheta||^2$.

\paragraph{Multilinear model}

Following~\cite{bayer:icd}, the derivation in this subsection makes the assumption that $\embu$ and $\embi$ do not share model parameters.
In particular for every scalar coordinate $\ttheta$ from the vector $\btheta$ of model parameters
\begin{align}
    \forall c \in C: \gradttheta \embu = 0 \quad \text{or} \quad \forall i \in I: \gradttheta \embi = 0 \label{eq:retrieval_grad}
\end{align}
To simplify notation, it is further assumed that the model is multilinear in the model parameters.
That means for each model parameter $\ttheta$ the model can be written in a linear form
\begin{align}
      \hyc{i} = g(i|c) + \ttheta \, \gradttheta \hyc{i}, \quad \gradgradttheta \hyc{i} = 0
\end{align}
Often, models are not just linear in scalars but linear in subvectors $\bttheta$ of the model parameters $\btheta$
\begin{align}
      \hyc{i} = g(i|c) + \langle \bttheta , \gradbttheta \hyc{i}\rangle, \quad \gradgradbttheta \hyc{i} = 0
\end{align}
For example, matrix factorization:
$    \hy(i|c) = \langle \bw_u, \bh_i \rangle$
is linear in $\bw_u$ or $\bh_i$.
Some other examples for multilinear models are factorization machines~\cite{rendle:fm}, or tensor factorization models such as PARAFAC~\cite{harshman:parafac} and Tucker Decomposition~\cite{tucker:td}.

\paragraph{Optimization}

With these prerequisites, efficient alternating least squares solvers can be derived.
Let $\bttheta_1, \bttheta_2, \ldots$ be a partition of the model parameters $\btheta$ such that the model is linear in each $\bttheta$.
From this follows that the optimal values for $\bttheta$ have a closed form solution that can be obtained by a linear regression solver.
An efficient computation of the least square solution needs to consider the Gramian trick.
This will be discussed next.
The solution can be derived by finding the root of $\gradbttheta L(\btheta,S)$.
A single iteration of Newton's method finds its solution:
\begin{align*}
    \bttheta^* = \bttheta - (\gradgradbttheta \tilde{L}(\btheta,\tilde{S}))^{-1} \gradbttheta \tilde{L}(\btheta, \tilde{S}) \label{eq:newton}
\end{align*}
The computation of the sufficient statistics $\gradgradbttheta \tilde{L}(\btheta,\tilde{S})$ and $\gradbttheta \tilde{L}(\btheta, \tilde{S})$ is now discussed for a vector of model parameters $\bttheta$ from the context side.
The equations for model parameters from the item side are analogously.
\begin{align*}
    \gradbttheta \tilde{L}(\btheta,\tilde{S}) &\stackrel{(*)}{=} \sum_{(c,i,\alpha, y) \in \tilde{S}} \alpha (\hyc{i} - y) \gradbttheta \hyc{i} + \alpha_0 \sum_{c \in C}  \embu G^I (\gradbttheta \embu)^t +  \lambda \bttheta \\
    \gradgradbttheta \tilde{L}(\btheta,\tilde{S}) &\stackrel{(**)}{=} \sum_{(c,i,\alpha, y) \in \tilde{S}} \alpha \gradbttheta \embu \otimes \gradbttheta \embu + \alpha_0 \sum_{c \in C}  (\gradbttheta \embu) G^I (\gradbttheta \embu)^t + \lambda I
\end{align*}
where (*) uses $\gradbttheta\hyc{i} = \langle \gradbttheta \embu, \embi \rangle$ and (**) uses of $\gradgradbttheta \hyc{i} = 0$.

From these equations follows the generic ALS algorithm from implicit data.
The algorithm iterates over parameters from the context side, computes the first and second derivative as outlined above and performs the Newton update step.
The same steps are repeated for parameters from the item side.
\begin{algorithm}[t]
  \caption{iALS-Pointwise for Multilinear Models}
  \label{alg:elementwise_als}
  \begin{algorithmic}[1]
        \Repeat
            \State $G^I \leftarrow \sum_{i \in I} \embi \otimes \embi$
            \For{$\bttheta \in \{\bttheta_1, \bttheta_2, \ldots\}$} \Comment{Iterate over parameters of the context side}
                \State $\gradbttheta \leftarrow \lambda \bttheta$
                \State $\gradgradbttheta \leftarrow \lambda I$
                \For{$c \in C$ where $\gradbttheta \embu \ne 0$}
                    \State $\gradbttheta \leftarrow \gradbttheta  + \alpha_0 \embu G^I (\gradbttheta \embu)^t$
                    \State $\gradgradbttheta \leftarrow \gradgradbttheta + \alpha_0 (\gradbttheta \embu) G^I (\gradbttheta \embu)^t$
                \EndFor
                \For{$(c,i,\alpha,y) \in \tilde{S}$ where $\gradbttheta \embu \ne 0$}
                    \State $\gradbttheta \leftarrow \gradbttheta + \alpha (\hyc{i} - y) \langle \gradbttheta \embu, \embi \rangle$
                    \State $\gradgradbttheta \leftarrow \gradgradbttheta + \alpha \gradbttheta \embu \otimes \gradbttheta \embu$
                \EndFor
                \State $\bttheta \leftarrow \bttheta - (\gradgradbttheta)^{-1} \gradbttheta$ 
            \EndFor
            \State{Perform a similar pass over parameters from the item side}
        \Until converged
  \end{algorithmic}
\end{algorithm}

If $\hy$ is a matrix factorization model, and if for $\bttheta$ a user embedding vector, $\bw_u$ or item embedding vector $\bh_i$ is chosen, then this is equivalent to the iALS algorithm proposed in~\cite{hu:ials}.
This algorithm has a complexity of $\O(|S|\,d^2 + (|C|+|I|)\,d^3)$ per epoch.
This is much more efficient than the complexity of $\O(|C|\,|I|\,d^2 + (|C|+|I|)\,d^3)$ of a naive ALS implementation.
The work of~\cite{hu:ials} was the first to provide this efficient training algorithm for learning matrix factorization from implicit feedback.
Later, \cite{hidasi:ialstensor} derived a variation of this algorithm for PARAFAC tensor factorization.

The derivation in this article applies to any multilinear model including matrix factorization, PARAFAC, Tucker Decomposition or factorization machines.
The coordinate descent version (i.e., choosing a scalar for $\btheta$) of this generalized algorithm was proposed in~\cite{bayer:icd}.
The alternating least squares version in this article is more general because any vector $\btheta$ can be chosen, including a vector of size one (i.e., a scalar) which would reduce to a coordinate descent algorithm.
Coordinate descent (CD) algorithms have a lower computation complexity than ALS solvers, e.g., $\O(|S|\,d + (|C|+|I|)\,d^2)$ for matrix factorization (see~\cite{bayer:icd} for details).
However, on modern hardware, vector operations as used by a ALS can be much more efficient than the scalar operations of CD, and with a careful implementations the higher theoretical complexity is not noticable.
Moreover, using vectors instead of scalars reduces synchronization points which is very important in distributed algorithms, so ALS methods can be overall more efficient in wall time than their CD equivalents.

\subsubsection{SGD Solver for General Models}

The previous section discussed multilinear models.
Now, following~\cite{krichene:isgd}, this is generalized to any dot-product model $\hyc{i} = \langle \embu, \embi \rangle$.
In particular, $\embu$ and $\embi$ can be any structure, including non-linear DNNs.
Such structures are commonly optimized by SGD algorithms that iterate over observed examples $S$.

Now, the loss in eq.~\eqref{eq:gramloss} is reformulated as a sum over training examples
\begin{align}
  \tilde{L}(\btheta,\tilde{S}) = \sum_{(c,i,\alpha,y) \in \tilde{S}}l(c,i,\alpha,y)
\end{align}
so that stochastic gradient descent can be applied.
To achieve this form, the Gramian term in the loss is rewritten as a sum over training examples:
\begin{align}
    \left\langle G^C, G^I \right\rangle 
&= \frac{1}{2}\left( \sum_{c \in C} \embu G^I \embu^t + \sum_{i \in I}  \embi G^C \embi^t\right) \\
&= \sum_{(c,i,\alpha,y) \in \tilde{S}} \frac{1}{2}\left( \frac{1}{|I_c|} \embu G^I \embu^t + \frac{1}{|C_i|} \embi G^C \embi^t \right)
\end{align}
Combining this with the loss on the positive observations, results in the final form of the elementwise loss
\begin{align}
  l(c,i,\alpha,y) = \alpha (\hyc{i} - y)^2 + \frac{\alpha_0}{2}\left( \frac{1}{|I_c|} \embu G^I \embu^t + \frac{1}{|C_i|} \embi G^C \embi^t\right)
\end{align}
Now, gradient descent can be applied to this equation.
For the purpose of learning $\btheta$,  $G^C$ and $G^I$ are treated as constants and replaced by estimates $\hat{G}^C$ and $\hat{G}^I$
\begin{multline}
  \gradbtheta l(c,i,\alpha,y) = 2\alpha (\hyc{i} - y) \gradbtheta \hyc{i} \\+ \alpha_0 \left( \frac{1}{|I_c|}  \embu \hat{G}^I (\gradbtheta\embu)^t + \frac{1}{|C_i|} \embi \hat{G}^C (\gradbtheta\embi)^t\right) \label{eq:sgd_gramian_loss}
\end{multline}
The Gramian estimates are updated by gradient descent as well.
A good Gramian estimate, $\hat{G}$, is close to the the true Gramian, $G$.
A reasonable objective to enforce closeness is the Frobenius norm $||\hat{G}-G||^2_F$.
To make this objective amendable to SGD, the Gramian loss is reformulated as a sum over positive training examples
\begin{align}
    \argmin_{\hat{G}^C}\left\lVert\hat{G}^C - G^C\right\rVert^2_F &= \argmin_{\hat{G}^C} \left\lVert\hat{G}^C - \sum_{c\in C} \embu \otimes \embu\right\rVert^2_F \\
    &= \argmin_{\hat{G}^C} \sum_{c\in C} \left\lVert\hat{G}^C -  |C| \embu \otimes \embu\right\rVert^2_F\\
    &= \argmin_{\hat{G}^C} \sum_{(c,i,\alpha,y) \in \tilde{S}} \frac{1}{|I_c|} \left\lVert\hat{G}^C -  |C| \embu \otimes \embu\right\rVert^2_F
\end{align}
which results in the gradient of the Gramian estimate:
\begin{align}
    &\grad_{\hat{G}^C} \left[ \sum_{(c,i,\alpha,y) \in \tilde{S}} \frac{1}{|I_c|} \left\lVert\hat{G}^C -  |C| \embu \otimes \embu\right\rVert^2_F \right] \\
    &= \sum_{(c,i,\alpha,y) \in \tilde{S}} \frac{1}{|I_c|} \left(\hat{G}^C -  |C| \embu \otimes \embu\right)
\end{align}
To summarize, when sampling a training example $(c,i,\alpha,y) \in \tilde{S}$, the update rule for the Gramian estimate is
\begin{align}
    \hat{G}^C &\leftarrow  \hat{G}^C - \eta  \frac{1}{|I_c|} \left(\hat{G}^C -  |C| \embu \otimes \embu \right) \\
    &= \hat{G}^C \left(1 - \frac{\eta}{|I_c|}\right) + \frac{\eta}{|I_c|}  |C| \embu \otimes \embu 
\end{align}
See \cite{krichene:isgd} for other estimation algorithms\footnote{
\cite{krichene:isgd} uses a different weight on each Gramian element which makes the derivation more natural for SGD algorithms.
This article uses the same Gramian definition as in eq~\eqref{eq:gramians} to make all results consistent.}.

Algorithm \ref{alg:elementwise_gravity} shows the final SGD procedure.

\begin{algorithm}[t]
  \caption{SGD-Pointwise with Gramian Trick}
  \label{alg:elementwise_gravity}
  \begin{algorithmic}[1]
        \Repeat
            \State sample $(c,i) \in S$
            \State $\btheta \leftarrow \btheta - \eta \gradbtheta l(c,i,\alpha,y)$ \Comment{See eq.~\eqref{eq:sgd_gramian_loss}} \label{line:elementwise_gravity_update}
            \State sample $(c,i) \in S$
            \State $\hat{G}^C \leftarrow  \hat{G}^C - \eta  \frac{1}{|I_c|} \left(\hat{G}^C -  |C| \embu \otimes \embu \right)$
            \State $\hat{G}^I \leftarrow  \hat{G}^I - \eta  \frac{1}{|C_i|} \left(\hat{G}^I -  |I| \embi \otimes \embi \right)$
        \Until converged
  \end{algorithmic}
\end{algorithm}

Comparing this to the vanilla SGD algorithm~\ref{alg:pointwise_sgd}, it can be seen that there is no negative sampling step necessary when using the Gramian trick.
All negatives are already considered in the update step of the positive item through the Gramian (see line~\ref{line:elementwise_gravity_update}).
In addition, there is the estimation step of the Gramians.

In some sense, negative sampling can be seen as an estimation of Gramians as well: instead of keeping track of an estimate of the Gramian as proposed here, negative sampling rebuilds the Gramian based on only a small sample of $m$ negatives.
This is a much less precise estimate of the Gramian than keeping a long-term estimate because the global Gramians are unlikely to change considerably for each step.

\subsection{Pairwise Square Loss}

Finally, efficient solvers for pairwise square losses are shortly discussed.
For keeping the derivation simple, it is assumed that $\alpha(c,i,j)=1$; a generalization to $\alpha(c,i,j) = \alpha_{c,i} \alpha_j$ is simple and follows the same pattern.
The pairwise square loss is defined as
\begin{align}
    L(\btheta, S) = \sum_{(c,i) \in S} \sum_{j \in I}[\hyc{i} - \hyc{j} - 1]^2 .
\end{align}
This can be reformulated to:
\begin{align}
    L(\btheta, S) =& \sum_{(c,i) \in S} \sum_{j \in I} [(\hyc{i}-1)^2 -2 (\hyc{i}-1)\hyc{j} + \hyc{j}^2] \\
    =& \sum_{(c,i) \in S} |I| (\hyc{i}-1)^2 + \langle G^C, G^I \rangle - 2 \sum_{(c,i) \in S}  (\hyc{i}-1) \langle \embu, \bz \rangle
\end{align}
with
\begin{align}
    G^C := \sum_{c \in C} |I_c|  \embu \otimes \embu, \quad
    G^I := \sum_{i \in I}  \embi \otimes \embi, \quad
    \bz := \sum_{i \in I} \embi
\end{align}
The first two terms in the loss are identical to the pointwise square loss in eq.~\eqref{eq:gramloss}: a pointwise loss over the positive observations, and the Gramian term.
The last term is an additional correction term for pairwise square loss, and its computation is in $\O(|S|\,d)$.
The algorithms (both iALS and SGD) introduced for pointwise loss can be extended to take the new term into account.
Instead of just storing the Gramian $G^I$, the vector $\bz$ has to be stored as well.

\cite{takasc:pairwiseials} first introduced an efficient solver for matrix factorization and pairwise square loss.
Their derivation is based on a gradient reformulation as previously invented for pointwise square loss by Hu et al.~\cite{hu:ials}.
The derivation above is more general and makes pairwise square loss applicable to a wider variety of models (multilinear models and general dot-product models) and optimization schemes (ALS, CD, SGD).

\section{Retrieval with Item Recommenders}
\label{sec:serving}

The typical application of an item recommender has to return the highest scoring items given a context $c$.
This section will first highlight the limitations of a naive brute-force implementation and then discuss efficient algorithms for dot product models.

\subsection{Limitations of Brute-Force Retrieval}

A naive implementation scores all the items, $I$, with $\hyc{i}$, sorts the scores and returns the highest ranked ones.
This is very costly, with a linear dependency on $|I|$.
This makes this method infeasible for online scoring for large or medium sized item catalogues because applications commonly expect the results in a few milliseconds within a user's request.
For moderately sized catalogues, the top items can be precomputed offline and stored for each context that could be queried.
However, the size of possible query context can be much larger than the set of training context $|C|$ and evaluating and storing all top lists becomes infeasible.
For example, for a sequential recommender, the context might be the $k$ most recent clicks of the user, the potential space of all context is $|I|^k$.
It is impossible to precompute the top scoring items for all of these $|I|^k$ context.
Offline computation becomes also a problem if the recommender system is trained online, e.g., a user's embedding might be updated online which would require recomputing the top scoring items.
To summarize, brute-force algorithms are only applicable if the number of items is small or for a static model if the number of context is small.
In the former case, brute-force scoring can be applied online, in the latter, brute-force scoring can be applied offline.

\subsection{Approximate Nearest Neighbor Search}
\label{sec:mips}

For dot product models, $\hyc{i}= \langle \embu, \embi \rangle$, the problem of finding the top scoring items is equivalent to the well studied field of \emph{maximum inner product search} or \emph{approximate nearest neighbor search}~\cite{wang:mips_survey}.
The nearest neighbor search problem is to find the closest entries in a database of vectors for a query vector.
For item recommendation, the context $\embu$ corresponds to the query and the database of vectors are the items $I$ represented by the embeddings $\embi$.
These problems are very well studied with efficient sublinear approximate algorithms that can be applied for the item recommendation problem.
Typically, solutions are based on narrowing down the set of possible nearest neighbors that needs to be evaluated, e.g., using partitioning algorithms such as trees \cite{yianilos:mips,muja:trees}, and efficient scoring using quantization~\cite{wang:mips_survey}.

These efficient sublinear algorithms make dot product models very attractive for large scale item recommendation.
If an application requires fast retrieval, dot product models are the best choice.

\subsection{Dynamic User Model}

When a user interacts with a recommender system, the system should be responsive and change the recommendation based on a user's feedback.
For example, after watching a video, the system should be able to make better recommendations taking into account this new information.
Depending on the type of scoring function, this might require to update the model itself.
For example, in a user-item matrix factorization model, the user embedding $\bw_u$ would need to be retrained in real time after every interaction -- for user-item matrix factorization a projection, eq.~\eqref{eq:newton} can be applied.
Other scoring functions avoid retraining by representing the user as a function of their past behavior.
For example, if the user is represented by the average item embedding in their history, the user embedding function $\embu$ is trivial to compute and can be done online at query time.
In both cases, the main model can be retrained occasionally offline to propagate all training data through the model.

\section{Conclusion}

This article introduced the item recommendation problem and discussed its unique challenges.
Several popular approaches to train recommender systems from implicit feedback data were presented.
An advantage of sampling based approaches is that they can be applied to most recommender models.
That makes them a popular choice for learning item recommenders.
Another advantage is that they can be adapted to retrieval metrics by rank based weighting and sampling schemes.
However, if the item catalogue is large, sampling based approaches are slow unless they use good samplers which is an open area of research.
For special cases the Gramian trick allows to learn item recommenders much more efficently, typically dropping the runtime dependency on the catalogue size $|I|$.
However, this trick is only applicable to square losses which can be less effective than a ranking loss such as weighted pairwise or softmax.
Dot product models that represent the context and the items by a $d$-dimensional embedding have useful properties throughout learning and retrieval.
For learning, their structure can be exploited for faster sampling, or the Gramian trick, and for retrieval it allows real-time retrieval in sublinear time through maximum inner product search algorithms.

As a conclusion, a few rules of thumb for applying recommender systems in practice are suggested.
Dot product models are a reasonable default choice due to their attractive retrieval properties.
The family of dot product models is broad including popular approaches such as matrix factorization but also more complex techniques such as deep neural networks including transformers, recurrent structures, or in general any model that extracts representations which are combined in the final layer with a dot product.
For learning, sampling based algorithms such as sampled softmax with a large number of samples $m$ are easy to implement and can result in good ranking quality.
If the item catalogue is very large, algorithms based on the Gramian trick can be a better choice, and implementations for models like matrix factorization are simple.
In any case, properly tuning and setting up models can be more important than switching to a more complex approach \cite{rendle:baselines,dacrema:reproducibility,rendle:ncf}.

\section*{Acknowledgements}
I would like to thank Nicolas Mayoraz and Li Zhang for helpful comments and suggestions.

\bibliographystyle{acm}

\end{document}